\documentclass[11pt]{article}

\usepackage[authoryear, sort&compress, round]{natbib}
\usepackage{graphicx}
\usepackage[small]{caption}
\usepackage{amsmath}
\usepackage{subcaption}
\usepackage[hyphens]{url}
\usepackage{hyperref}
\hypersetup{breaklinks=true}
\usepackage{cite}

\usepackage{xcolor}

\usepackage{geometry}
\usepackage{titlesec}
\usepackage{authblk}

\titleformat{\section}{\large\bfseries}{\thesection}{1em}{}

\title{Memory, Consciousness and Large Language Model}

\author[1]{Jitang Li\footnote{Equally contribute} \thanks{lijitang@suda.edu.cn}}
\author[2]{Jinzheng Li$^*$ \thanks{lijinzheng22@gmail.com}}

\affil[1]{Department of Philosophy, Soochow University, Suzhou, Jiangsu 215006, China}
\affil[2]{Independent Researcher}
\date{} 

\begin{document}
\maketitle
\begin{abstract}
    {{With the development in cognitive science and Large Language Models (LLMs), increasing connections have come to light between these two distinct fields. Building upon these connections, we propose a conjecture suggesting the existence of a duality between LLMs and Tulving's theory of memory. We identify a potential correspondence between Tulving's synergistic ecphory model (SEM) of retrieval and the emergent abilities observed in LLMs, serving as supporting evidence for our conjecture. Furthermore, we speculate that consciousness may be considered a form of emergent ability based on this duality. We also discuss how other theories of consciousness intersect with our research.}}
\end{abstract}
\section{Introduction}
{{
Consciousness, one of the oldest mysteries, has intrigued humanity for millennia. However, for a long time, the exploration of consciousness remained within armchair philosophy. Serious scientific research into consciousness began in the last century, and it has since evolved into a complex and intricate field of study \citep{seth2022theories}. On the other hand, despite our lack of understanding regarding the underlying principles, LLM demonstrates astonishing capabilities across a wide range of tasks \citep{wei2022emergent}. These two topics originate from distinct research areas, but there is a growing willingness to discuss them together. Such discussion leads to an unavoidable question: Can artificial intelligence(AI) like LLMs become conscious? This question has been discussed in some recent papers like \citep{butlin2023consciousness,ledoux2023consciousness}. 

Instead of addressing this question directly, let's consider it from another perspective: What do we, as conscious human beings, have in common with LLMs?  The answer that immediately comes to mind is memory. The study of memories in LLMs is crucial for solving issues like catastrophic forgetting and hallucination. Moreover, there exists a profound connection between memory and consciousness, as elucidated by Tulving's research in the last century \citep{tulving1985memory}. During our research on memories in LLM and in Tulving's theory of memory, we have uncovered a series of intriguing coincidences between these two seemingly disparate subjects. Thus we propose a bold conjecture suggesting the presence of a duality between Tulving's memory theory and various memories in LLM. Expanding upon this duality, we further identify a potential correspondence between Tulving’s synergistic ecphory model (SEM) of retrieval and the emergent abilities observed in LLMs. This perspective offers a novel approach to comprehending emergent abilities and in-context learning. This theory is consistent with several existing experimental observations, including \citep{lu2023emergent,min2022rethinking,wang2023label,chan2022data,wei2022emergent,brown2020language}. The duality finally leads us to attribute the relationship between memory and consciousness in Tulving's theory as an emergent ability.

The structure of this paper is outlined as follows: In Section \ref{sec2}, we introduce Tulving's theory of memory and consciousness. In Section \ref{sec:dual}, we propose a conjecture suggesting the existence of a duality between memories in LLM and those in Tulving's theory. We provide a detailed explanation of how each type of memory corresponds between these two distinct areas. In Section \ref{SEM/EA}, building upon this duality, we establish a potential correspondence between the synergistic ecphory model of retrieval and emergent ability. Additionally, we present supporting evidence for our argument. In Section \ref{CasEA}, we speculate that consciousness is a form of emergent ability, drawing upon our previous arguments and other corroborating evidence. In Section \ref{dis}, we further discuss our consciousness theory in detail. Finally, the conclusion and further study are given in Section \ref{conc}.
}}
\section{Tulving's Theory of Memory and Consciousness \label{sec2}}
Various methods exist for categorizing different types of memories. In 1963, Melton distinguished between three essential steps in the learning and memory process: encoding, storage, and retrieval \citep{melton1963implications}. Memories can also be classified based on the duration for which information remains accessible: sensory memory, short-term memory, and long-term memory \citep{atkinson1968human}.

However, in this paper, we adhere to Tulving's theory of memory, which categorizes memories based on the properties of their content. Tulving proposed this distinction in his work \citep{tulving1972episodic,tulving1985memory}, identifying three distinct types of memories: procedural, semantic, and episodic. Procedural memory pertains to how tasks and actions are performed. Semantic memory involves the storage of symbolically representable knowledge about the world. Episodic memory plays a role in remembering personally experienced events. Psychologists have continued to explore better ways to delineate these distinctions, but Tulving's framework remains a cornerstone in the field \citep{de2022rethinking}. It's important to note that such a distinction represents a choice of scientific perspective, which is neither inherently right nor wrong but rather convenient. It has proven to be highly appropriate for studying memory. 
\begin{table}[h]
    \centering
    \begin{tabular}{ccc}
    \hline
         Memory system& & Consciousness  \\
    \hline\\
     Episodic&{\huge$\Longleftrightarrow$}  &Autonoetic\\
     ${\huge\downarrow}$& &${\huge \downarrow}$\\
     Semantic&{\huge$\Longleftrightarrow$} & Noetic\\
     ${\huge\downarrow}$& &${\huge \downarrow}$\\
     Procedural&{\huge$\Longleftrightarrow$}  & Anoetic\\
     \hline
    \end{tabular}
    \caption{A schematic diagram of the relation between memory systems and varieties of consciousness.}
    \label{tab:schematic}
\end{table}

In Tulving's theory, each of the three memory systems is characterized by a different form of consciousness: anoetic (non-knowing), noetic (knowing), and autonoetic (self-knowing). Table.\ref{tab:schematic} illustrates these relationships. A classic example of procedural memory is riding a bicycle. When you ride a bicycle, it doesn't trigger memories of your previous experiences, and you don't need to consciously think about how to ride. Semantic memory pertains to general knowledge, such as remembering that \textit{The capital of France is Paris.} Episodic memory encompasses personally experienced events. 

\section{Duality between LLM and Tulving's Theory of Memory \label{sec:dual}}
{{ Duality is a very powerful tool that is widely used in the fields of physics and math. {\textit{Fundamentally, duality gives two different points of view of looking at the same object. Many things have two different points of view and in principle they are all dualities }\citep{atiyah2007duality}. In a broad sense, this is also an approach from a scientific perspective. }It excels in the construction of new theories based on some existing theories. In this paper, we propose a conjecture suggesting the existence of a duality between LLMs and Tulving’s theory of memory. We now proceed to explain how and why this duality is plausible.

We begin by examining the memory systems within these two theories. In our previous discussion, we outlined the various types of memory in Tulving's theory. The question now is whether we can identify corresponding memories in LLM. First, it's important to note that LLM is renowned for its extensive knowledge reservoir. LLMs acquire knowledge through either pre-training or fine-tuning processes, implicitly storing it within their parameters. This type of knowledge aligns with the definition of semantic memory. When it comes to procedural memory, LLM does not have procedural memories in the traditional sense, like riding a bike or playing basketball. However, there are analogous behaviors. For instance, LLM is capable of recognizing the need to insert a line break "$\backslash n$" at the end of each paragraph. Additionally, through fine-tuning, it is possible to introduce catchphrases into LLM's responses, such as instructing it to append "Meow" to the end of each sentence.

Identifying the corresponding memory for episodic memory in LLM is of paramount importance. For episodic memory, an essential aspect has been discussed in Tulving's work. Tulving\citep{tulving1985memory,budson2022consciousness} defined episodic memory as the set of processes that enable us to mentally time-travel and re-experience past moments. These processes involve the initial intake of information from our sensory stores and working memory, followed by the creation of a mental representation of a specific moment in time. In the realm of Natural Language Processing (NLP), we interpret such a description to refer to "time-series" information. We can no longer retain episodic memories in the same manner as we can with the other two types of memory. This is primarily because the current architecture of LLM cannot store time-series data within its parameters. In other words, when you store a date by fine-tuning, it serves merely as a numerical value for the LLMs, rather than representing a genuine recollection of the past.

Coincidentally, there is another element that plays the role of episodic memory perfectly: the input context. Input context refers to the tokens or words that the model considers when making next-token predictions. Due to the autoregressive sampling design of the LLM and the positional embedding in the Transformer, the chromatic (time-series) information is provided to the LLM, satisfying our need for episodic memory. Thus, we have successfully established a correspondence between the memories in the LLM and Tulving's theory as Table.(\ref{tab:schematic2}) shows. 

{{Similar attempts to establish a correspondence between memory in psychology and memory in LLM have been made in many articles about LLM agents \citep{weng2023prompt}.  However, their arguments are based on the current limitation in context length, and therefore they require an external vector database for long-term memory. It is another "bitter lesson"\citep{sutton2019bitter} happening right now.
}}

\begin{table}[h]
    \centering
    \begin{tabular}{ccccc}
    \hline
     LLM Memory& &    Memory system  \\
    \hline\\
     Input context&{\huge$\Longleftrightarrow$}  &Episodic\\
     ${\huge\downarrow}$& &${\huge \downarrow}$\\
    Memory by pre-train/finetune &{\huge$\Longleftrightarrow$}  &Semantic\\
     ${\huge\downarrow}$& &${\huge \downarrow}$\\
     Memory by pre-train/finetune&{\huge$\Longleftrightarrow$}  &Procedural\\
     \hline
    \end{tabular}
    \caption{A schematic diagram of the relation between memory systems from Tulving's theory and different memories in LLMs.}
    \label{tab:schematic2}
\end{table}

Formally proving this duality will be exceedingly challenging. However, once this duality is established, additional correspondences will emerge spontaneously. For instance, Tulving's papers\citep{tulving1985memory,tulving2002episodic} offer case studies of amnesic patients, some of whom exhibit behavior closely resembling that of a LLM with limited context length. Offering additional correspondence with evidence like this can significantly bolster our conjecture. Two more pieces of supporting evidence are provided in Appendix \ref{appA} and the following section.

}}

 {{
\section{Synergistic-Ecphory-Model/Emergent-Ability Correspondence \label{SEM/EA}}

The emergent abilities of LLMs have been extensively studied and discussed in previous research \citep{wei2022emergent,lu2023emergent,schaeffer2023emergent}.
Generally, emergent abilities of large language models as abilities that are
not present in smaller-scale models but are present in large-scale models. In essence, the philosophy behind the emergent abilities of LLM is mainly the so-called “more is different” by famous physicist Philip Anderson \citep{anderson1972more}. Unfortunately, the precise mechanisms underlying these emergent abilities remain unknown.

In the last section, we tried to establish a duality between LLM and Tulving's memory theory. Based on this duality, we can anticipate that the emergent abilities observed in LLM will align with corresponding theories within Tulving's memory framework. Coincidently, such a theory does indeed exist, known as the synergistic ecphory model (SEM) of retrieval.
\begin{figure}[t]
    \centering
    \begin{subfigure}{0.8\textwidth}
    \includegraphics[width=\textwidth]{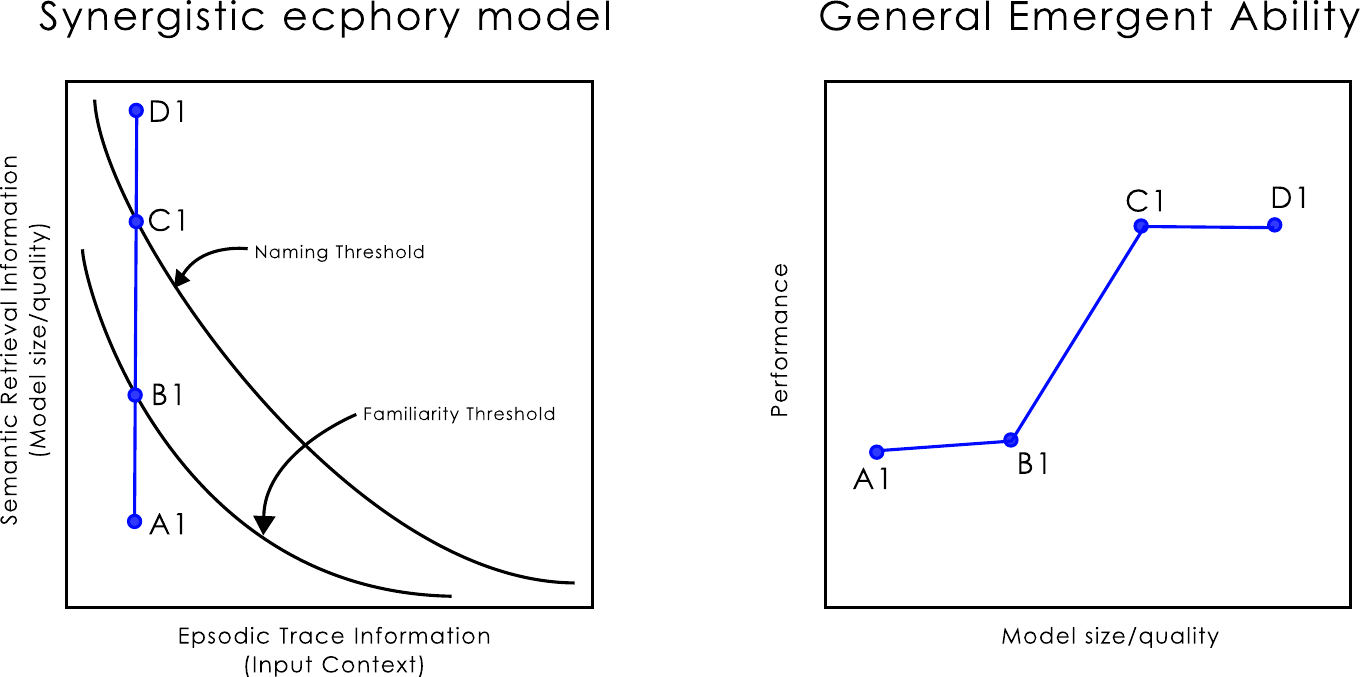}
    \caption{General emergent ability}
    \label{fig:SEM}
    \end{subfigure}
    \\
    \vspace{0.5cm}
    \begin{subfigure}{0.8\textwidth}
    \includegraphics[width=\textwidth]{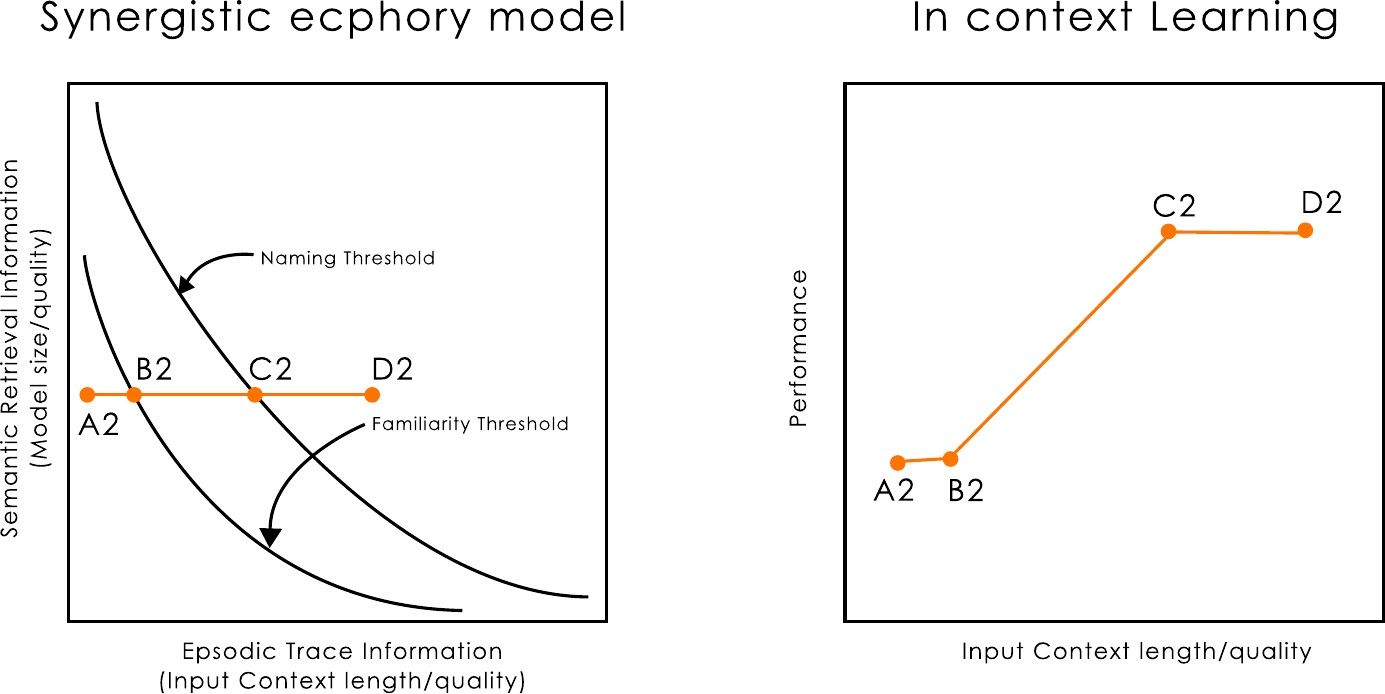}
    \caption{In context learning}
    \label{fig:SEM2}
    \end{subfigure}
    \caption{Schematic diagrams of synergistic ecphory model of retrieval corresponding to (a) general emergent ability and (b) in-context learning.}
\end{figure}

In Tulving's paper\citep{tulving1985memory,tulving1982synergistic}, he introduces the SEM to explain how knowledge about past events can be recovered from the episodic system and the semantic system. A comprehensive description of this model can be found in \citep{tulving1982synergistic}. Here we just provide a concise summary of it. A process of ecphory is the process of how appropriate information is extracted from the cue and brought into interaction with the stored episodic information. The product of a successful act of ecphory is referred to as ecphoric information. A schematic description of the model is shown in the left panel of Fig.(\ref{fig:SEM}). The horizontal axis of the coordinate system represents episodic trace information and the vertical axis represents semantic retrieval information. The two axes of the coordinate system represent both the quantity and quality of trace information and retrieval information. The two curved lines in the diagram represent two conversion thresholds, the lower for familiarity judgments of the kind made in the recognition task, and the upper for the production of the name of the retrieved item-event, as required in the recall task.  The two conversion thresholds divide the total space of ecphoric information into three regions. Region 1: The region below the familiarity threshold consists of ensembles of ecphoric information that are insufficient for recognition. Region 2: The region between the two thresholds represents bundles of ecphoric information that contain and provide sufficient evidence for making positive familiarity judgments but insufficient evidence for the construction of the name of the original item-event. Region 3: The region above the naming threshold represents ecphoric information that is sufficient for the production of the name of the target event. The shape of conversion thresholds in the diagram is arbitrary, but it must satisfy two features: The thresholds are asymptotic with the two coordinate axes. The naming threshold must be above the familiarity threshold. 

Next, we establish a correspondence between the SEM and emergent abilities as follows: Ecphoric information corresponds to emergent abilities. This is possible because abilities are not fundamentally different from information for a neural network. Episodic trace information corresponds to the input context, and semantic retrieval information corresponds to the knowledge stored through pre-training and fine-tuning, as previously discussed. Therefore, the horizontal axis represents the length and quality of the input context, while the vertical axis represents the model size, amount of training data, training duration, and more.

The blue line in Fig.(\ref{fig:SEM}) has the same horizontal axis value, indicating that we are using identical trace information.  In the context of LLM, this implies the application of a specific prompting strategy. Then we increase the vertical axis value from A1 to D1, signifying an increase in either the model's size or the amount of training data.  As we transition from A1 to B1, we remain within Region 1, where the ecphoric information (emergent ability) remains unattained. Progressing from B1 to C1, we enter Region 2 and attain some familiarity with ecphoric information, although it remains insufficient to fully realize this ability. Consequently, we observe an improvement in performance as the model size increases. Transitioning from C1 to D1 places us in Region 3, where the ecphoric information (emergent abilities) is fully achieved, rendering further improvement unlikely. We draw the performance curve on the right panel of Fig.(\ref{fig:SEM}), which aligns with performance curves of most emergent abilities. Similarly, we observe a correspondence between the SEM and in-context learning in LLM, as illustrated in  Fig.(\ref{fig:SEM2}). A large model contains a large amount of semantic information, which means only a little episodic information is needed to reach the familiarity threshold. It explains why LLM can function as a few-shot learner. The same deduction can be applied to other few-shot prompting methods, such as the chain of thought.

Here we provide some existing experimental observations that match our theory regarding this correspondence.
\begin{itemize}
    \item \citep{wei2022emergent,brown2020language} The SEM successfully explained the shape of the performance curve for emergent ability and in-context learning. Most importantly, it accounted for both factors simultaneously.
    \item \citep{lu2023emergent} It has been demonstrated that emergent performance does not manifest in the absence of in-context learning in this work. This phenomenon can be perfectly explained by the SEM that the ecphoric process requires both trace information (input context) and retrieval information (model size/quality).
    \item \citet{min2022rethinking} It has been observed that LLMs do not learn new tasks during test time. Their analysis shows that the model may ignore the task defined by the demonstrations and instead use prior from pertaining. This phenomenon can be better understood in the view of the ecphory process that the abilities are actually "recalled" from the model.
\end{itemize}   Furthermore, we have noticed that several other papers align with our conjecture: \citep{wang2023label,chan2022data}.

It's important to emphasize that we are not attempting to assert that "Emergent ability is merely the retrieval of memory." Our intention is simply to illustrate the connection between the SEM and emergent ability. It is also conceivable that "Memory retrieval is a form of emergent ability." However, this topic lies beyond the scope of our current discussion.
}}

\section{Consciousness as an Emergent Ability\label{CasEA}}
{{So far, all the emergent abilities we have observed have originated from LLMs with substantial model sizes and extensive training data. However, if our theory regarding the correlation between the SEM of retrieval and the emergent abilities of LLMs is accurate, we should also anticipate witnessing emergent abilities generated by relatively smaller models with significantly extended context lengths, as illustrated in Fig.(\ref{fig:longcontextEA}). We firmly believe that such emergent abilities do indeed exist, and one plausible candidate among them is consciousness. (In this section, consciousness refers to the widely discussed general consciousness, closely related to Tulving's concept of autonoetic consciousness.) 
\begin{figure}[t]
    \centering
    \includegraphics[width=0.8\textwidth]{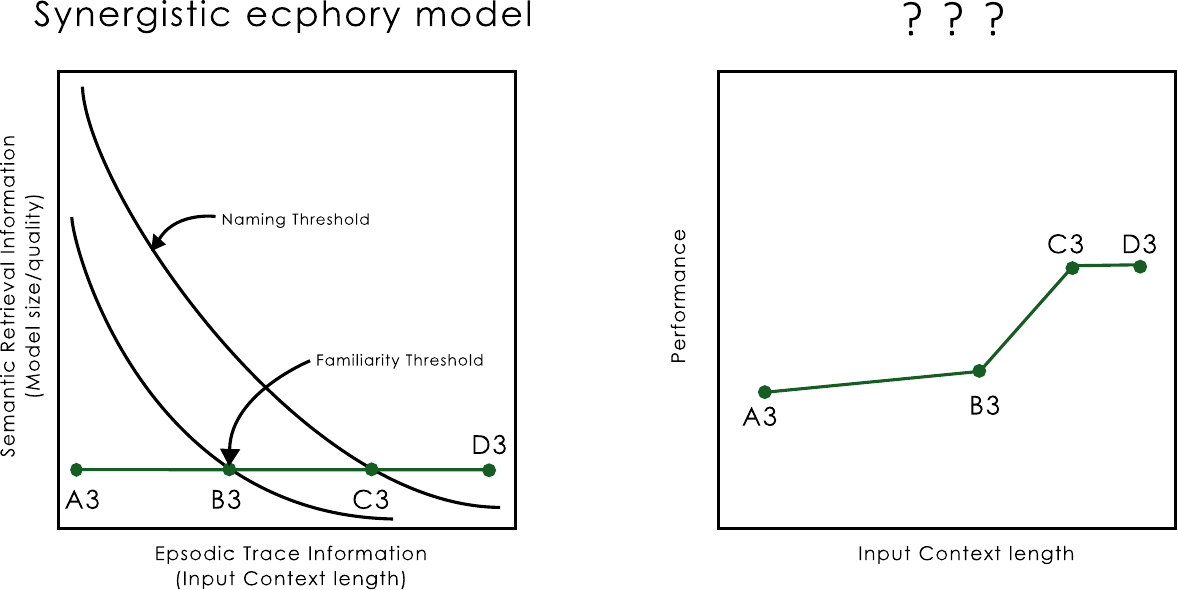}
    \caption{Schematic diagrams of synergistic ecphory model of retrieval corresponding to some unknown emergence ability.}
    \label{fig:longcontextEA}
\end{figure}

Here are some reasons to support this statement. First, as we discussed in section \ref{sec:dual}, each of the three memory systems is characterized by a
different kind of consciousness. However, Tulving did not make a clear statement of how different consciousnesses are connected to different memories. Based on our analysis in the previous section, we believe that it is very likely that consciousness will appear in the way of emergent abilities once a sufficient reservoir of long-term memory is established. Secondly, psychologists have long been aware that infants do not exhibit self-awareness until they reach approximately two years of age \citep{kagan1984nature,lewis1995shame}. The emergence of an infant's consciousness remains unclear, but it is reasonable to speculate that their level of consciousness develops in conjunction with their memory capacity. A baby's ability to become self-aware depends on having sufficient memory and other cognitive abilities. Lastly, consciousness is not an exceedingly advanced function in nature. Many animals exhibit consciousness even though their level of intelligence is quite low \citep{low2012cambridge}. Thus, a smaller model should not be a problem for the existence of consciousness.

Until now, there is no evidence suggesting the development of a potential emergent ability arising from an extended input context. Here, we will present three reasons to explain why this emergence has not been observed yet. Additionally, this discussion outlines the future research direction in this field.
\begin{itemize}
    \item The current context length remains insufficient for the development of an emergent ability. One characteristic of emergent abilities is that \textit{performance is near-random until a certain critical threshold of scale is reached, after which performance increases to substantially above random.}\citep{wei2022emergent} However, we do not yet know what this threshold should be, which implies it could be a very large number. Nevertheless, due to the quadratic complexity of transformers, most of the current Large Language Models (LLMs) have a maximum context length of 4k to 16k tokens.  A few LLMs have now achieved approximately 100k tokens, but even this extended length may still not suffice to generate the emergent ability. One potential solution to this problem is to employ a linear complexity architecture, such as RWKV\citep{peng2023rwkv} or Mamba\citep{gu2023mamba}, to attain an infinite context length. In fact, for these State Space Models (SSMs), one notable property is that their context memory typically decays exponentially over time. Coincidently, short-term memory also generally exhibits exponential decay over time in psychological studies \citep{peterson1959short}.
\item Input context and training data are distinct entities. Both CLS, as discussed in Appendix \ref{appA}, and SEM, as described in Section \ref{SEM/EA}, directly or indirectly suggest a strong connection between episodic memory and semantic memory. However, current LLMs are typically trained on random information from the internet, which is often unrelated to the input context. A data-dependent decay method, as introduced in some current SSMs\citep{gu2023mamba}, appears to alleviate this issue by removing unrelated context memories while retaining the strongly related ones. However, in an ideal scenario, all training data should either fully or partially align with the input context, similar to the CLS approach.
\item Current examination metrics cannot accurately assess the presence of emergent abilities over an extended context. As indicated by reference \citep{schaeffer2023emergent}, the assessment of emergent abilities is greatly affected by the choice of metrics and testing methods.  Certain tests, like the one outlined in \citep{longcontectclaude}, are typically geared toward evaluating the accuracy of the task. Nevertheless, when it comes to assessing ambiguous emergent abilities, such as consciousness, innovative testing approaches may be required.

\end{itemize}
}}
\section{Discussion  \label{dis}}
The theoretical framework of "consciousness as an emergent ability" primarily relies on the duality between LLM and Tulving's theory of memory. It is commonly believed that consciousness exists in a unitary state, with one notable exception being Tulving's tripartite taxonomy of "autonoetic," "noetic," and "anoetic" consciousness. Given the broad scope of Tulving's theory, LeDoux and Birch invited ten leading experts to approach questions regarding the consciousness of AI and animals within Tulving's framework \citep{ledoux2023consciousness}. Their discussion illustrates that Tulving's theory of memory consciousness is well-suited for addressing the issue of consciousness in humans, animals, and AI. It prompts us to think further about Tulving's theory of memory and leads us to explore the duality between LLM and Tulving's theory.

In Tulving's theory, he argues that consciousness is best discussed at the level of memory. This point is particularly crucial.  There are various competing theories of consciousness: The Integrated Information Theory (IIT) begins with the phenomenal experience of consciousness and posits that consciousness can be understood in terms of 'cause-effect power' associated with irreducible maxima of integrated information generated by physical systems. The Global Neuronal Workspace Theory (GNWT) suggests that sensory information gains access to consciousness when it is 'broadcast' within the neuronal workspace. However, these theories of consciousness cannot overcome the hard problem of consciousness. The issue arises from the attempt to investigate consciousness directly through the lens of information, which primarily deals with the easy problem, rather than addressing the hard problem. By contrast, Tulving's theory of consciousness is rooted in the memory system, the measurements of which are well-known to us, rather than being directly tied to information. Thus, for Tulving's theory, the hard problem of consciousness is less prominent. Although GNWT emphasizes the role of memory and attention, it needs further explanation regarding the so-called process of ignition, which will be revisited from the perspective of information.

The theory of "consciousness as an emergent ability" differs from the theory of consciousness as an "emergent property" of matter. The former falls within the realm of cognitive science, while the latter pertains to the philosophy of mind. When we discuss consciousness as an "emergent property" of matter, we are attempting to address the "mind-body" problem, a topic within metaphysics. Even if one were to substitute "matter" with "life + special neurobiological features", it would still be a metaphysical perspective \citep{feinberg2020phenomenal}. Cognitive science is an interdisciplinary field that explores the realms of mind and intelligence, encompassing philosophy, psychology, artificial intelligence, neuroscience, linguistics, and anthropology. Its foundation lies in the mind-computer analogy, which has evolved into a complex 3-way analogy involving the mind, the brain, and computers \citep{sep-cognitive-science}. As a scientific phenomenological theory, "consciousness as an emergent ability" encompasses both human consciousness as an emergent ability and the consciousness of LLM as an emergent ability. This theory possesses a relatively comprehensive theoretical framework, comprising empirical phenomena, mathematical relationships, phenomenological models, and fundamental theories from a bottom-up perspective. It encompasses the differentiation of consciousness within three memory systems, the duality between LLM and Tulving's theory of memory, the correspondence between the SEM and the emergent ability of LLM, and the concept of consciousness as an emergent ability. Moreover, this theory can elucidate conscious phenomena in humans and predict the emergence of conscious phenomena in LLM.

One crucial point to emphasize is the significance of the concept of time series. In Kant's philosophical framework, there exists a transcendental nature of time: time serves as a form of \textit{Anschauungsformen}, i.e., a form of the intuition of our self and our inner state, playing a fundamental role in determining inner intuition to represent the temporal sequence of all objects of senses, just as space does for intuition. Without spatial intuitive forms, we would be unable to perceive the spatial arrangement of the external world. Similarly, without time as a form of inner sense, we would struggle to grasp the chronological sequence of events. Consequently, it becomes apparent that the input context functioning as episodic memory is essential within this understanding of time series. This concept serves as the foundation of the duality.

\section{Conclusion and Future Study \label{conc}}
The understanding of human cognition has consistently served as a source of inspiration for AI research. Meanwhile, scholars explore cognitive science by drawing analogies between the human mind, the brain, and computers. Based on Tulving's memory theory, we try to refine these analogies into a duality framework. Guided by this duality relationship, we uncover a potential correspondence between SEM and the emergence ability.
Consequently, we propose a theory of consciousness as an emergent ability, applicable not only to human consciousness but also to potential LLM consciousness.  The study of human consciousness as an emergent ability and the study of AI consciousness as an emergent ability mutually bolster and inspire one another, contributing to the resolution of consciousness-related challenges.
The research presented in this article predominantly adopts a phenomenological approach, which currently represents one of the most effective methods for investigating the enigmatic workings of the human brain and LLMs.

For future studies, while we have presented a series of pieces of supporting evidence based on the work of others, conducting additional direct experiments is of utmost importance at this juncture. One potential experiment is to quantitatively establish the familiarity threshold curve and naming threshold for a specific emergent ability within LLM. If such a curve indeed exists, it would serve as compelling evidence for our conjecture. As we discussed in section \ref{CasEA}, SSMs exhibit a strong alignment with our theory. This architecture is very likely to be the one that attains emergent abilities through the extension of context length, rather than the scaling of model size. We should pay more attention to SSMs. Furthermore, this work is also crucial for the field of AI safety. If our hypotheses are correct, restricting the context length will be a highly efficient way to avoid some potential risks while maintaining performance on LLMs.

If further research substantiates our conjecture of the existence of a duality between LLMs and Tulving's theory of memory, we propose naming this duality the \textit{Tulving-LLM duality} in memory of the late and esteemed experimental psychologist and cognitive neuroscientist, Endel Tulving, who passed away a few months ago.

\section*{Acknowledgments}
We would like to thank Yuyan Liu for discussion. The research of Jitang Li was supported in part by National Social Science Fund of China (grant No.23BZX112), Humanities and Social Sciences Program of Ministry of Education (grant No.20YJA720005).  

\appendix
\section{Appendix A: Complementary Learning Systems (CLS) and LLM \label{appA} }

The Complementary Learning Systems (CLS) theory in cognitive neuroscience proposes that the brain has two interconnected systems for learning and memory, each with distinct functions. The Hippocampal System, centered in hippocampus, is crucial for creating new, episodic memories, allowing for quick learning and temporary information storage. The Neocortical System, involving the neocortex, excels in the slow integration and long-term storage of information, effectively consolidating knowledge over time. These systems operate in tandem: the hippocampal system rapidly acquires new information, which is then transferred to the neocortical system for permanent storage and integration with existing knowledge. This synergy explains the quick learning of new information but also the need for time to achieve deep understanding and retention.

CLS is consistent with the memory system in LLM. We can describe the entire process in the language of LLM as follows: Fresh information enters the memory system initially as the input context. Information and knowledge are subsequently consolidated from the input context to avoid data contamination. The consolidated results are then stored within parameters through fine-tuning. This establishes correspondences between the input context and the hippocampal system, as well as between the neural network in LLM and the neocortical system.

For a long time, there has been a prevailing belief that the hippocampal system is primarily responsible for short-term memory storage. However, recent studies have challenged this notion by demonstrating its capacity to store long-term memory as well. Several new theories have emerged to explain this phenomenon \citep{yonelinas2019contextual, sun2023organizing}. Correspondingly, it is essential to consider the inclusion of long-term memory within the input context. These long-term episodic memories play a pivotal role in the synergistic ecphory process discussed in section \ref{SEM/EA}.

\bibliography{main}

\begin{thebibliography}{33}
\providecommand{\natexlab}[1]{#1}
\providecommand{\url}[1]{\texttt{#1}}
\expandafter\ifx\csname urlstyle\endcsname\relax
  \providecommand{\doi}[1]{doi: #1}\else
  \providecommand{\doi}{doi: \begingroup \urlstyle{rm}\Url}\fi

\bibitem[lon()]{longcontectclaude}
Long context prompting for claude 2.1.
\newblock \url{https://www.anthropic.com/index/claude-2-1-prompting}.
\newblock Accessed: 2023-12-12.

\bibitem[Anderson(1972)]{anderson1972more}
P.~W. Anderson.
\newblock More is different: Broken symmetry and the nature of the hierarchical structure of science.
\newblock \emph{Science}, 177\penalty0 (4047):\penalty0 393--396, 1972.

\bibitem[Atiyah(2007)]{atiyah2007duality}
M.~Atiyah.
\newblock Duality in mathematics and physics.
\newblock \emph{Confer{\`e}ncies FME}, 5:\penalty0 2007--2008, 2007.

\bibitem[Atkinson and Shiffrin(1968)]{atkinson1968human}
R.~C. Atkinson and R.~M. Shiffrin.
\newblock Human memory: A proposed system and its control processes.
\newblock In \emph{Psychology of learning and motivation}, volume~2, pages 89--195. Elsevier, 1968.

\bibitem[Brown et~al.(2020)Brown, Mann, Ryder, Subbiah, Kaplan, Dhariwal, Neelakantan, Shyam, Sastry, Askell, Agarwal, Herbert-Voss, Krueger, Henighan, Child, Ramesh, Ziegler, Wu, Winter, Hesse, Chen, Sigler, Litwin, Gray, Chess, Clark, Berner, McCandlish, Radford, Sutskever, and Amodei]{brown2020language}
T.~B. Brown, B.~Mann, N.~Ryder, M.~Subbiah, J.~Kaplan, P.~Dhariwal, A.~Neelakantan, P.~Shyam, G.~Sastry, A.~Askell, S.~Agarwal, A.~Herbert-Voss, G.~Krueger, T.~Henighan, R.~Child, A.~Ramesh, D.~M. Ziegler, J.~Wu, C.~Winter, C.~Hesse, M.~Chen, E.~Sigler, M.~Litwin, S.~Gray, B.~Chess, J.~Clark, C.~Berner, S.~McCandlish, A.~Radford, I.~Sutskever, and D.~Amodei.
\newblock Language models are few-shot learners, 2020.

\bibitem[Budson et~al.(2022)Budson, Richman, and Kensinger]{budson2022consciousness}
A.~E. Budson, K.~A. Richman, and E.~A. Kensinger.
\newblock Consciousness as a memory system.
\newblock \emph{Cognitive and Behavioral Neurology}, 35\penalty0 (4):\penalty0 263, 2022.

\bibitem[Butlin et~al.(2023)Butlin, Long, Elmoznino, Bengio, Birch, Constant, Deane, Fleming, Frith, Ji, Kanai, Klein, Lindsay, Michel, Mudrik, Peters, Schwitzgebel, Simon, and VanRullen]{butlin2023consciousness}
P.~Butlin, R.~Long, E.~Elmoznino, Y.~Bengio, J.~Birch, A.~Constant, G.~Deane, S.~M. Fleming, C.~Frith, X.~Ji, R.~Kanai, C.~Klein, G.~Lindsay, M.~Michel, L.~Mudrik, M.~A.~K. Peters, E.~Schwitzgebel, J.~Simon, and R.~VanRullen.
\newblock Consciousness in artificial intelligence: Insights from the science of consciousness, 2023.

\bibitem[Chan et~al.(2022)Chan, Santoro, Lampinen, Wang, Singh, Richemond, McClelland, and Hill]{chan2022data}
S.~C.~Y. Chan, A.~Santoro, A.~K. Lampinen, J.~X. Wang, A.~Singh, P.~H. Richemond, J.~McClelland, and F.~Hill.
\newblock Data distributional properties drive emergent in-context learning in transformers, 2022.

\bibitem[De~Brigard et~al.(2022)De~Brigard, Umanath, and Irish]{de2022rethinking}
F.~De~Brigard, S.~Umanath, and M.~Irish.
\newblock Rethinking the distinction between episodic and semantic memory: Insights from the past, present, and future.
\newblock \emph{Memory \& Cognition}, 50\penalty0 (3):\penalty0 459--463, 2022.

\bibitem[Feinberg and Mallatt(2020)]{feinberg2020phenomenal}
T.~E. Feinberg and J.~Mallatt.
\newblock Phenomenal consciousness and emergence: eliminating the explanatory gap.
\newblock \emph{Frontiers in psychology}, 11:\penalty0 1041, 2020.

\bibitem[Gu and Dao(2023)]{gu2023mamba}
A.~Gu and T.~Dao.
\newblock Mamba: Linear-time sequence modeling with selective state spaces, 2023.

\bibitem[Kagan(1984)]{kagan1984nature}
J.~Kagan.
\newblock \emph{The nature of the child.}
\newblock Basic Books, 1984.

\bibitem[LeDoux et~al.(2023)LeDoux, Birch, Andrews, Clayton, Daw, Frith, Lau, Peters, Schneider, Seth, et~al.]{ledoux2023consciousness}
J.~LeDoux, J.~Birch, K.~Andrews, N.~S. Clayton, N.~D. Daw, C.~Frith, H.~Lau, M.~A. Peters, S.~Schneider, A.~Seth, et~al.
\newblock Consciousness beyond the human case.
\newblock \emph{Current Biology}, 33\penalty0 (16):\penalty0 R832--R840, 2023.

\bibitem[Lewis(1995)]{lewis1995shame}
M.~Lewis.
\newblock \emph{Shame: The exposed self}.
\newblock Simon and Schuster, 1995.

\bibitem[Low et~al.(2012)Low, Panksepp, Reiss, Edelman, Van~Swinderen, and Koch]{low2012cambridge}
P.~Low, J.~Panksepp, D.~Reiss, D.~Edelman, B.~Van~Swinderen, and C.~Koch.
\newblock The cambridge declaration on consciousness.
\newblock In \emph{Francis crick memorial conference, Cambridge, England}, pages 1--2, 2012.

\bibitem[Lu et~al.(2023)Lu, Bigoulaeva, Sachdeva, Madabushi, and Gurevych]{lu2023emergent}
S.~Lu, I.~Bigoulaeva, R.~Sachdeva, H.~T. Madabushi, and I.~Gurevych.
\newblock Are emergent abilities in large language models just in-context learning?
\newblock \emph{arXiv preprint arXiv:2309.01809}, 2023.

\bibitem[Melton(1963)]{melton1963implications}
A.~W. Melton.
\newblock Implications of short-term memory for a general theory of memory.
\newblock \emph{Journal of verbal Learning and verbal Behavior}, 2\penalty0 (1):\penalty0 1--21, 1963.

\bibitem[Min et~al.(2022)Min, Lyu, Holtzman, Artetxe, Lewis, Hajishirzi, and Zettlemoyer]{min2022rethinking}
S.~Min, X.~Lyu, A.~Holtzman, M.~Artetxe, M.~Lewis, H.~Hajishirzi, and L.~Zettlemoyer.
\newblock Rethinking the role of demonstrations: What makes in-context learning work?, 2022.

\bibitem[Peng et~al.(2023)Peng, Alcaide, Anthony, Albalak, Arcadinho, Biderman, Cao, Cheng, Chung, Grella, GV, He, Hou, Lin, Kazienko, Kocon, Kong, Koptyra, Lau, Mantri, Mom, Saito, Song, Tang, Wang, Wind, Wozniak, Zhang, Zhang, Zhao, Zhou, Zhou, Zhu, and Zhu]{peng2023rwkv}
B.~Peng, E.~Alcaide, Q.~Anthony, A.~Albalak, S.~Arcadinho, S.~Biderman, H.~Cao, X.~Cheng, M.~Chung, M.~Grella, K.~K. GV, X.~He, H.~Hou, J.~Lin, P.~Kazienko, J.~Kocon, J.~Kong, B.~Koptyra, H.~Lau, K.~S.~I. Mantri, F.~Mom, A.~Saito, G.~Song, X.~Tang, B.~Wang, J.~S. Wind, S.~Wozniak, R.~Zhang, Z.~Zhang, Q.~Zhao, P.~Zhou, Q.~Zhou, J.~Zhu, and R.-J. Zhu.
\newblock Rwkv: Reinventing rnns for the transformer era, 2023.

\bibitem[Peterson and Peterson(1959)]{peterson1959short}
L.~Peterson and M.~J. Peterson.
\newblock Short-term retention of individual verbal items.
\newblock \emph{Journal of experimental psychology}, 58\penalty0 (3):\penalty0 193, 1959.

\bibitem[Schaeffer et~al.(2023)Schaeffer, Miranda, and Koyejo]{schaeffer2023emergent}
R.~Schaeffer, B.~Miranda, and S.~Koyejo.
\newblock Are emergent abilities of large language models a mirage?
\newblock \emph{arXiv preprint arXiv:2304.15004}, 2023.

\bibitem[Seth and Bayne(2022)]{seth2022theories}
A.~K. Seth and T.~Bayne.
\newblock Theories of consciousness.
\newblock \emph{Nature Reviews Neuroscience}, 23\penalty0 (7):\penalty0 439--452, 2022.

\bibitem[Sun et~al.(2023)Sun, Advani, Spruston, Saxe, and Fitzgerald]{sun2023organizing}
W.~Sun, M.~Advani, N.~Spruston, A.~Saxe, and J.~E. Fitzgerald.
\newblock Organizing memories for generalization in complementary learning systems.
\newblock \emph{Nature neuroscience}, 26\penalty0 (8):\penalty0 1438--1448, 2023.

\bibitem[Sutton(2019)]{sutton2019bitter}
R.~Sutton.
\newblock The bitter lesson.
\newblock \emph{Incomplete Ideas (blog)}, 13\penalty0 (1), 2019.

\bibitem[Thagard(2023)]{sep-cognitive-science}
P.~Thagard.
\newblock {Cognitive Science}.
\newblock In E.~N. Zalta and U.~Nodelman, editors, \emph{The {Stanford} Encyclopedia of Philosophy}. Metaphysics Research Lab, Stanford University, {W}inter 2023 edition, 2023.

\bibitem[Tulving(1982)]{tulving1982synergistic}
E.~Tulving.
\newblock Synergistic ecphory in recall and recognition.
\newblock \emph{Canadian Journal of Psychology/Revue canadienne de psychologie}, 36\penalty0 (2):\penalty0 130, 1982.

\bibitem[Tulving(1985)]{tulving1985memory}
E.~Tulving.
\newblock Memory and consciousness.
\newblock \emph{Canadian Psychology/Psychologie canadienne}, 26\penalty0 (1):\penalty0 1, 1985.

\bibitem[Tulving(2002)]{tulving2002episodic}
E.~Tulving.
\newblock Episodic memory: From mind to brain.
\newblock \emph{Annual review of psychology}, 53\penalty0 (1):\penalty0 1--25, 2002.

\bibitem[Tulving et~al.(1972)]{tulving1972episodic}
E.~Tulving et~al.
\newblock Episodic and semantic memory.
\newblock \emph{Organization of memory}, 1\penalty0 (381-403):\penalty0 1, 1972.

\bibitem[Wang et~al.(2023)Wang, Li, Dai, Chen, Zhou, Meng, Zhou, and Sun]{wang2023label}
L.~Wang, L.~Li, D.~Dai, D.~Chen, H.~Zhou, F.~Meng, J.~Zhou, and X.~Sun.
\newblock Label words are anchors: An information flow perspective for understanding in-context learning, 2023.

\bibitem[Wei et~al.(2022)Wei, Tay, Bommasani, Raffel, Zoph, Borgeaud, Yogatama, Bosma, Zhou, Metzler, et~al.]{wei2022emergent}
J.~Wei, Y.~Tay, R.~Bommasani, C.~Raffel, B.~Zoph, S.~Borgeaud, D.~Yogatama, M.~Bosma, D.~Zhou, D.~Metzler, et~al.
\newblock Emergent abilities of large language models.
\newblock \emph{arXiv preprint arXiv:2206.07682}, 2022.

\bibitem[Weng(2023)]{weng2023prompt}
L.~Weng.
\newblock Llm-powered autonomous agents.
\newblock \emph{lilianweng.github.io}, Jun 2023.
\newblock URL \url{https://lilianweng.github.io/posts/2023-06-23-agent/}.

\bibitem[Yonelinas et~al.(2019)Yonelinas, Ranganath, Ekstrom, and Wiltgen]{yonelinas2019contextual}
A.~P. Yonelinas, C.~Ranganath, A.~D. Ekstrom, and B.~J. Wiltgen.
\newblock A contextual binding theory of episodic memory: systems consolidation reconsidered.
\newblock \emph{Nature Reviews Neuroscience}, 20\penalty0 (6):\penalty0 364--375, 2019.

\end{thebibliography}

\end{document}